\documentclass[conference, 10pt, onecolumn]{IEEEtran}
\usepackage{amsmath}
\usepackage[dvips]{graphicx}
\usepackage{amsmath}
\usepackage{amssymb}
\usepackage{epsfig}
\usepackage{graphicx}
\usepackage{dsfont}
\usepackage{float}
\usepackage{algorithm2e}
\usepackage{xfrac}
\usepackage{color}
\usepackage{psfrag}
\usepackage{amsthm}
\usepackage{subfigure}

\DeclareMathAlphabet{\bm}{OML}{cmr}{bx}{it}
\DeclareMathAlphabet{\mathsf}{OT1}{cmss}{m}{n}
\DeclareMathAlphabet{\bs}{OT1}{cmss}{bx}{it}

\newcommand{\rs}{\mathrm}
\newcommand{\mc}{\mathcal}

\newcommand{\cluster}{\text{CH}}

\newtheorem{remark}{Remark}

\title{Exploiting Interference for Efficient Distributed Computation in Cluster-based Wireless Sensor Networks}

\author{
\IEEEauthorblockN{Steffen Limmer\IEEEauthorrefmark{1},
		S\l awomir
                Sta\'nczak\IEEEauthorrefmark{1}\IEEEauthorrefmark{2},
                Mario Goldenbaum\IEEEauthorrefmark{2},
		Renato L. G. Cavalcante \IEEEauthorrefmark{1}
		}
		\IEEEauthorblockA{
	\IEEEauthorrefmark{1}
		Fraunhofer Institute for Telecommunications, Heinrich Hertz Institute,
		Einsteinufer 37,  10587 Berlin, Germany,\\
    \IEEEauthorrefmark{2}
    Fachgebiet Informationstheorie und theoretische Informationstechnik,\\
    Technische Universit\"at Berlin, Einsteinufer 25, 10587 Berlin, Germany\\		
	 Email: \{steffen.limmer, slawomir.stanczak, renato.cavalcante\}@hhi.fraunhofer.de; mario.goldenbaum@tu-berlin.de
    }

\thanks{Disclaimer: This work has been accepted for publication in the IEEE Proc. 1st IEEE Global Conference on Signal and Information Processing (GlobalSIP '13). Copyright with IEEE. Personal use of this material is permitted. However, permission to reprint/republish this material for advertising or promotional purposes or
for creating new collective works for resale or redistribution to servers or lists, or to reuse
any copyrighted component of this work in other works must be obtained from the IEEE.
This material is presented to ensure timely dissemination of scholarly and technical work.
Copyright and all rights therein are retained by authors or by other copyright holders. All
persons copying this information are expected to adhere to the terms and constraints
invoked by each author’s copyright. In most cases, these works may not be reposted
without the explicit permission of the copyright holder. For more details, see the IEEE
Copyright Policy.}
}

\begin{document}
\IEEEoverridecommandlockouts
%
\maketitle

\begin{abstract}
  This invited paper presents some novel ideas on how to enhance the performance of
  consensus algorithms in distributed wireless sensor
  networks, when communication costs are considered. Of particular interest
  are consensus algorithms that exploit the broadcast property of the wireless
  channel to boost the performance in terms of convergence speeds. To this end, we
  propose a novel clustering based consensus algorithm that exploits
  interference for computation, while reducing the energy consumption in the
  network. The resulting optimization problem is a semidefinite program, which
  can be solved offline prior to system startup. 
\end{abstract}
%
%

\section{Introduction}

\label{sec:intro}

In many wireless (sensor) applications, nodes cooperate for some common
goal. One example is the localization of an acoustic source using a number of
geographically distributed microphones that are equipped with wireless
communication capabilities. In fire alarm networks, for instance, a number of
wireless sensor nodes may be used to monitor maximum and average temperature
values. In such applications, the goal is therefore not to share local
measurements among network nodes but rather to compute one or multiple
functions of these measurements (e.g., the maximum function or a weighted
sum). The functions to be computed depend on the targeted application.

A key observation is that it is in general not necessary to exchange raw
sensor measurements in order to compute a function thereof. It is further
known that the broadcast property of the wireless channel can be beneficially
exploited when the task is to compute or estimate function values at sensor
nodes. In particular, the information-theoretic analysis in \cite{NazerGastparInfoTh} shows how to encode messages for linear multiple-access channels to
enhance the rate at which linear functions can be reliably computed at the
receiver.

A different approach can be found in \cite{Goldenbaum:Stanczak:11a} where an
\emph{analog} computation scheme is proposed which 1) is able to efficiently
compute \emph{non-linear functions} over many multiple-access fading channels
and 2) is robust against the lack of synchronization between different
signals.\footnote{Only a coarse frame synchronization is needed.} The idea was
used in \cite{ZhGoStYu12} as a building block of a cluster-based average
algorithm to improve the convergence speed of gossip-based
algorithms for average consensus \cite{BoGhPrSh06}. Reference \cite{GBS12ngf}
extended the approach of \cite{ZhGoStYu12} to incorporate non-linear functions
($f$-consensus algorithms).

In this paper, we consider the approach of \cite{ZhGoStYu12} but take into
account different energy costs imposed on sensor nodes by different
distant-dependent path losses. We neglect the energy consumption for feedback
and channel estimation, which however is relatively low under certain
conditions \cite{GS10cfvs}. Given the setup, the objective is to cluster
sensor nodes and then activate the clusters in such a way as to minimize the
overall energy consumption for transmission. In each cluster, we neglect the
impact of noise and assume that the average of sensor measurements within the
cluster is computed using the CoMAC (Computation over Multiple-Access Channel)
scheme of \cite{Goldenbaum:Stanczak:11a} (see also Figure $1$). A consequence
of this is that the energy consumption can be reduced by decreasing the size
of the clusters; since this deteriorates the convergence speed and rate, there
is an inherent trade-off between the energy consumption and the convergence
behavior. This paper studies this trade-off by numerically solving a suitably
formulated clustering and activation optimization problem. Different points on
the trade-off curve are achieved by choosing different weights for a
regularization term that takes into account the energy costs.
The paper is organized as follows: We first introduce the system model and the
cost model. This is followed by the problem statement. In Section
\ref{sec:clustering}, we reformulate the original problem of maximizing the
convergence rate to take into account the energy consumption for
transmission. This problem is solved using some standard optimization
tools. Section \ref{sec:simulations} presents some simulations, while the
paper is completed by some conclusions and open problems.

\section{System Model}
\label{sec:sys}

\subsection{Network Model}
We consider a clustered wireless sensor network consisting of $N\in
\mathbb{N}$ (sensor) nodes that are grouped in $C$ clusters. Let
$\mc{N}=\{1,\dotsc,N\}$ and let $\mc{C}_i\subseteq\mc{N}$ be the index set of
nodes belonging to cluster $i\in\mc{C}:=\{1,\hdots,C\}$. Throughout the paper,
we use $\cluster_i\in\mc{C}_i$ to refer to the cluster head of the $i$th
cluster. Note that in general, $\mc{C}_i\cap\mc{C}_j\neq\emptyset$ for any
$i,j\in\mc{C}$ and $\cup_{i=1}^C\mc{C}_i=\mc{N}$.
\begin{remark}
  Note that $C$ denotes the number of possible clusters that are used in our
  optimization. $C$ may be very large and, in extreme case, is equal to
  $2^N-1$, the cardinality of the power set of $\mc{N}$ excluding the empty
  set. However, while operating, the network may activate only a subset of
  this set as a result of the optimization process.
\end{remark}

The sensor nodes are distributed over some geographical area and we use
$\bm{X}=[\bm{x}_1,\hdots,\bm{x}_N] \in \mathbb{R}^{2 \times N}$ to denote the
position matrix. Accordingly, $\bm{x}_n\in\mathbb{R}^2$ contains the
coordinate of sensor node $n$ with respect to some reference point. The vector
of squared distances from node $i$ to all nodes is $\bm{d}_i= \{\lVert
\bm{x}_i - \bm{x}_j \rVert^2\}_{j=1,\hdots,N}$ and we group these vectors in a
matrix $\bm{D}=[\bm{d}_1,\hdots,\bm{d}_N]$.

\subsection{Time Model}
To model the temporal behavior of cluster activation, the asynchronous time
model in \cite{BoGhPrSh06} is adopted. Furthermore we assume that the cluster
heads wake up according to a rate $\mu_i \in \mathbb{R}_+$ Poisson process,
where $\mu_i$ is chosen such that only a single cluster head is activated
within a certain time window (with high probability).

\subsection{Communication Model}
\begin{figure}
\label{fig:cluster1}
\centering \subfigure[Step 1: Function computation.]{
  \psfrag{x1}{$\bm{x}_1=\rs{CH}_i$} \psfrag{x2}{$\bm{x}_2$} \psfrag{x3}{$\bm{x}_3$}
  \psfrag{x4}{$\bm{x}_4$} \psfrag{x5}{$\bm{x}_5$} \psfrag{x6}{$\bm{x}_6$}
  \psfrag{txmax}{$d(\bm{x}_5,\bm{x}_1)$}
  \includegraphics[width=0.25\linewidth]{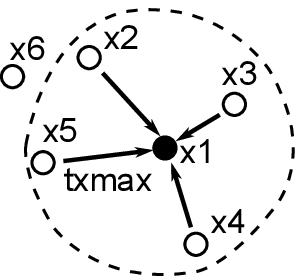}
} \subfigure[Step 2: Broadcast result.]{ \psfrag{x1}{$\bm{x}_1=\rs{CH}_i$}
  \psfrag{x2}{$\bm{x}_2$} \psfrag{x3}{$\bm{x}_3$} \psfrag{x4}{$\bm{x}_4$}
  \psfrag{x5}{$\bm{x}_5$} \psfrag{x6}{$\bm{x}_6$}
  \psfrag{txmax}{$d(\bm{x}_1,\bm{x}_5)$}
  \includegraphics[width=0.25\linewidth]{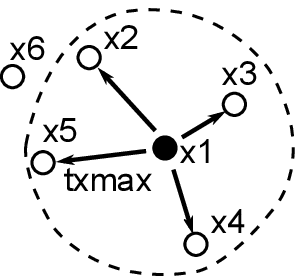}
}
\caption{Two-step approach to average consensus}
\end{figure}
Denote the reading of the $k$th sensor at time $t\in\mathbb{Z}_{+}$ by $y_k(t) \in \mc{Y} \subset \mathbb{R}$, $k \in
1,\hdots,N$, and the initial network state by $\bm{y}(0)\in \mc{Y}^N$.  If
cluster $\mc{C}_i$ is active at time $t\in\mathbb{Z}_{+}$, the received signal
$r_{\rs{CH}_i}$ at time $t$ by cluster head $i$ (node $1$ in Fig. $1$), is
given by
\begin{align}
r_{\rs{CH}_i}(t) = \sum_{k \in \mc{C}_i \backslash {\rs{CH}_i}} h_{ik}(t) s_k(y_k(t)) + n_i(t),
\end{align}
where $h_{ik}(t)$, $n_i(t)$ and $s_k(\cdot)$ denote the corresponding channel
coefficient from node $k$ to cluster head, receiver-side noise and transmit
signal of node $k$, respectively. This setting will be referred to as
$\emph{noisy MAC}$. Assume, nodes can estimate the channel to the cluster head
by some wake up pilot symbols. Then, to compute the average within cluster
$\mc{C}_i$, every node needs to invert it's own channel, which yields transmit
signals of the form
\begin{align}
s_k(y_k(t))=\frac{1}{h_{ik}(t)}y_k(t)
\end{align}
and simplifies the received signal to
\begin{align}
\label{equ:noisy_mac}
r_{\rs{CH}_i}(t) = \sum_{k \in \mc{C}_i \backslash {\rs{CH}_i}} y_k(t) + n_i(t).
\end{align}
Hence, the channel inversion removes the impact of the channel. \emph{In what
  follows, we assume an idealized noiseless setting; the noisy case remains
  an open problem for potential future research.}

Considering \eqref{equ:noisy_mac} and neglecting the noise term, we can write
the time evolution of the state vector in a compact form. To this end, let
$\bm{y}(t)=[y_1(t),\hdots,y_N(t)]$ be the state vector at time
$t\in\mathbb{Z}_+$ and let $p_i\in[0,1],i\in\mc{C}$, be the probability with
which cluster $i$ is activated. Then, $\bm{y}(t)$ evolves over time as
\begin{align}
  \bm{y}(t+1)=\bm{W}(t)\bm{y}(t),t\in\mathbb{Z}_+,
\end{align}
where for every $t\in \mathbb{Z}_+$, matrix
$\bm{W}(t):=[\bm{w}_1,\hdots,\bm{w}_N]\in \mathbb{R}^{N \times N}$ is
independently and randomly chosen according to the probability mass
function/vector
\begin{equation}
\label{eq:pmv}
\bm{p}=[p_1,\dotsc,p_C]\in[0,1]^C 
\end{equation}
from the set $\{\bm{W}^{(i)}\}_{i=1}^{C}$ of weight matrices defined by
\cite{ZhGoStYu12}
\begin{align}
W_{j,k}^{(i)} := \begin{cases}
1, & \text{if}\ j\notin \mc{C}_i,\ k=j \\
\frac{1}{N_i}, & \text{if}\ j,k\in\mc{C}_i \\
0, & \text{else.}
\end{cases}
\end{align}

\begin{remark}
  To compute functions different from arithmetic or weighted mean, the
  consensus step can be adapted from weight matrices $\bm{W}^{(i)}$ to
  functionals $f: \mc{Y}^N \mapsto \mathbb{R}$. It has been shown in
  \cite{GoBoSt13} that the resulting function space computable with the
  proposed scheme is essentially the space of nomographic functions, which
  contains also functions such as geometric mean as a special case.
\end{remark}
\begin{remark}
  For an analysis of the noisy setting, we refer the interested reader to
  \cite{CaSt13}.
\end{remark}

\section{Cost Model}
\label{sec:cost_model}

We employ the energy cost model of \cite{HeChBa00}, which, for a transmission
of a $k$-bit message over a distance $d_{i,j}$ from node $i$ to node $j$, is given by
\begin{align}
E_{\rs{tot}} = E_{\rs{tx}} + E_{\rs{rx}} = k E_{\rs{elec}} + \epsilon_{\rs{amp}} k d_{i,j}^2 + k E_{\rs{elec}}.
\end{align}
Notice that the circuitry for tx and rx is assumed to be identical for all
nodes. \emph{In what follows, we neglect the steady energy consumption
  ($E_{\rs{elec}}$) and consider only dynamic energy consumption caused by
  amplifiers in the tx circuitry averaged over some time period.} We however
point out that the results of this paper can be easily extended to capture the
steady energy consumption, which may strongly influence the optimization
results if the sensor nodes can be switched off after computing a function
value. In such cases, it may be beneficial to prefer larger clusters when the
steady energy consumption increases. \emph{Moreover, for brevity, we neglect
  the receiver-side energy consumption for reception, which could be easily
  incorporated as well.}

For an activated cluster $\mc{C}_i$ with respective weight matrix
$\bm{W}^{(i)}$ and corresponding cluster head $\rs{CH}_i \in \mc{C}_i$, we
have to differentiate between the energy cost associated with Step 1
$\bm{c}_{\rs{fc}}^{(i)}$ and the energy cost imposed by Step 2
$\bm{c}_{\rs{bc}}^{(i)}$ (see Figure $1$). In the first case,
the energy cost results from the transmission of sensor nodes to the $i$th
cluster head and is therefore given by
\begin{align}
\bm{c}_{\rs{fc}}^{(i)} = \bm{d}_{\rs{CH}_i} \odot \mathbf{1}_{\mc{C}_i}
\end{align}
where $\mathbf{1}_{\mc{C}_i}$ is the vector-valued indicator function of the
set $\mc{C}_i$ defined as follows: if $i\in\mc{C}_i$, then the $i$th element
of $\mathbf{1}_{\mc{C}_i}$ is $1$; otherwise it is zero.
Accordingly, the energy cost associated with Step 2 
is determined by the maximum distance between the cluster head and cluster
nodes, i.e. we have
\begin{align}
\bm{c}_{\rs{bc}}^{(i)} = \lVert \bm{d}_{\rs{CH}_i} \odot \mathbf{1}_{\mc{C}_i}\rVert_\infty \cdot\bm{e}_{\rs{CH}_i}
\end{align}
where $\bm{e}_{\rs{CH}_i}\in\mathbb{R}^N$ is a unit vector with a $1$ at the
$\rs{CH}_i$th position and zeros otherwise.

\begin{remark} Choosing an outer node as cluster head results in higher
  communication costs (as distances and therefore costs will be larger) even
  if the respective weight matrices are identical.
\end{remark}

\section{Problem Statement}
\label{sec:clustering}

Following the derivation in \cite{ZhGoStYu12,BoGhPrSh06}, improving the
convergence rate for a predefined clustered WSN can be achieved by optimizing
the cluster activation probabilities $\bm{p}=[p_1,\dotsc,p_C]$, corresponding
to weight matrices $\bm{W}^{(i)},i=1\cdots C$. The problem of optimizing the
convergence rate can therefore be formulated as the problem of minimizing the
second largest eigenvalue
\begin{equation}
\label{eq:2eigenvalue}
[0,1]^C\mapsto [0,1]:\xi(\bm{p}):=\lambda_2(\bm{W}(\bm{p}))
\end{equation}
of the stochastic weight matrix $\bm{W}:=\bm{W}(\bm{p})=\sum_i
p_i\bm{W}^{(i)}$ (see \cite{ZhGoStYu12,BoGhPrSh06}). Taking into account the
constraints, we arrive at the following optimization problem:
\begin{align}
\label{equ:prob_only}
\underset{\bm{p} \in [0;1]^C}{\operatorname{min}} \ & \xi(\bm{p}) \\
\text{s.t.} \ & \bm{W} - \bm{J} \preceq \xi(\bm{p}) \bm{I}_N \nonumber \\ 
& \bm{W}=\sum_{i=1}^{C} p_i \bm{W}^{(i)} \nonumber \\
& \sum_{i=1}^{C} p_i = 1 \nonumber 
\end{align}
where $N\bm{J}=\mathbf{1}\mathbf{1}^T\in\mathbb{R}^{N\times N}$ is the all-one
matrix. Note that the first constraint ensures that the second eigenvalue is
smaller than or equal to the largest eigenvalue of $\bm{W}$, which is $1$. An
upper bound for the resulting mean squared error can be given with
$\boldsymbol{\varepsilon}(t):=\bm{y}(t)-\overline{\bm{y}}(0)$ by
\cite{ZhGoStYu12,BoGhPrSh06}
\begin{align*}
  \mathbb{E} \left\{ \boldsymbol{\varepsilon}(t)^T \boldsymbol{\varepsilon}(t) \right\} & \leq \xi(\bm{p}) \boldsymbol{\varepsilon}(t-1)^T \boldsymbol{\varepsilon}(t-1) 
   \leq \xi(\bm{p})^t \boldsymbol{\varepsilon}(0)^T \boldsymbol{\varepsilon}(0)
\end{align*}

\section{Joint Energy/Convergence Optimization}
\label{sec:optimization}

Now we are in a position to extend the optimization problem in
(\ref{equ:prob_only}) to incorporate the additional energy cost. The problem
of interest - called the joint convergence/expected lifetime optimization
problem - takes the form
\begin{align}
\label{equ:prob_clus}
\underset{\bm{p} \in [0;1]^C}{\operatorname{min}} \ & \xi(\bm{p}) + \alpha \lVert \bm{c}(\bm{p}) \rVert_1 \\
\text{s.t.} \ &\bm{W} - \bm{J} \preceq \xi(\bm{p}) \bm{I}_N \nonumber \\
& \bm{W}=\sum_{i=1}^{M} p_i \bm{W}^{(i)} \nonumber \\
& \sum_{i=1}^{C} p_i = 1 \nonumber \\
& \bm{c}(\bm{p}) = \sum_{i=1}^C p_i ( \bm{c}_{\rs{fc}}^{(i)} + \bm{c}_{\rs{bc}}^{(i)}) \nonumber.
\end{align}
In addition, we assume that $\xi(\bm{p})\leq 1-\epsilon$ for some sufficiently
small $\epsilon>0$. This ensures that the resulting clusters are connected and
therefore the algorithm converges. Furthermore, we point out that the
associated constraint set $\{\bm{p}\in[0,1]^{C}:\xi(\bm{p})\leq 1-\epsilon\}$
is convex for any choice of $\epsilon$. This is because the largest eigenvalue
of positive semidefinite symmetric matrices is a convex function of the matrix entries
which depend linearly on $\bm{p}$. A practical problem is the choice of
$\epsilon$ which cannot be too large since otherwise the problem is
infeasible. 

We emphasize that the set of all possible weight matrices (i.e. clusterings)
$\{\bm{W}^{(i)}\}_{i=1,\hdots,C}$ comprises $C=(N-1)N$ elements (each node can
be cluster head of a cluster consisting of $\{2,\hdots,N\}$ nodes), which
might result in optimization problems that are very expensive from a
computational perspective. However, the set of weight matrices can be reduced if
we restrict the number of nodes per cluster. In addition, we can exclude
candidate clusters with an outer node as cluster head because these choices
correspond to identical consensus steps only at a higher cost, i.e
$\bm{W}^{(i)}=\bm{W}^{(j)}$, with $\lVert \bm{c}^{(i)} \rVert_1 \geq \lVert
\bm{c}^{(j)} \rVert_1$ for some $\mc{C}_i=\mc{C}_j$, $\rs{CH}_i \neq
\rs{CH}_j$.

\section{Simulation Results}
\label{sec:simulations}

In this Section we present simulation results for the proposed algorithm
\eqref{equ:prob_clus} using different predefined cluster sizes $\lvert
\mc{C}_i \rvert \subseteq \{2,\hdots,30\}$. The simulated WSN consists of
$N=30$ nodes that are placed uniformly at random in a $50\times50$ square and
to solve the optimiziation problem \eqref{equ:prob_clus} we use CVX
\cite{GrBo12}. To ensure convergence of the resulting consensus scheme, the
optimization variable $\xi(\bm{p})$ is bounded by $\xi(\bm{p}) \leq
1-\epsilon$. However, this bound must not be set too small, otherwise the
optimization problem will not be feasible. We found $\epsilon=10^{-2}$ to
yield good results.

To evaluate the resulting consensus scheme consisting of weight matrices
$\bm{W}^{(i)}$, respective costs $\bm{c}^{(i)}$ and activation probabilities
$\bm{p}$, we simulate a physical application, where the nodes are deployed to
monitor temperature values. The initial network state $\bm{y}(0)$ is drawn
uniformly from $[0 \ ^\circ\text{C},30 \ ^\circ\text{C}]^N$. We monitor the
error defined by $\frac{\lVert \boldsymbol{\varepsilon}(t) \rVert_2^2}{\lVert
  \bm{y}(0) \rVert_2^2}$ and the consumed energy for a certain realization of
the sequence of activated clusters. The results are averaged over $10^3$ runs.
For the simulations, we also assume that the nodes have expert knowledge of
the current estimation error to terminate communication, and therefore also
dynamic energy consumption, once the error falls below a threshold of
$10^{-1}$.

\begin{figure}[htb]
\label{fig:resultsnew1}
 \psfrag{Mean Estimation Error}{\textcolor{blue}{\small{Mean Estimation Error}}}
 \psfrag{Mean Consumed Energy}[t][c]{\textcolor{red}{\small{Mean Consumed Energy}}}
 \psfrag{Number t of Iterations}{\small{Number $t$ of Iterations}}
 \psfrag{0.0}{\small{$\alpha=0$}}
 \psfrag{4e-05              }{\small{$\alpha=4\cdot 10^{-5}$}}
 \psfrag{8e-05}{\small{$\alpha=8\cdot 10^{-5}$}}
 \psfrag{0}{\small{$0$}}
 \psfrag{0.05}{\small{$0.05$}}
  \psfrag{0.1}{\small{$0.1$}}
 \psfrag{0.15}{\small{$0.15$}}
 \psfrag{0.2}{\small{$0.2$}}
 \psfrag{0.25}{\small{$0.25$}}
 \psfrag{0.3}{\small{$0.3$}}
 \psfrag{0.35}{\small{$0.35$}}
 \psfrag{0.4}{\small{$0.4$}}
 \psfrag{10}{\small{$10$}}
 \psfrag{20}{\small{$20$}}
 \psfrag{30}{\small{$30$}}
 \psfrag{40}{\small{$40$}}
 \psfrag{12}{\small{$12$}}
 \psfrag{2000}{\small{$2000$}}
 \psfrag{4000}{\small{$4000$}}
 \psfrag{6000}{\small{$6000$}}
 \psfrag{8000}{\small{$8000$}}
 \psfrag{10000}{\small{$10000$}}
 \psfrag{12000}{\small{$12000$}}
 \psfrag{14000}{\small{$14000$}}
 \psfrag{}{}
 \centering
  \includegraphics[width=8cm]{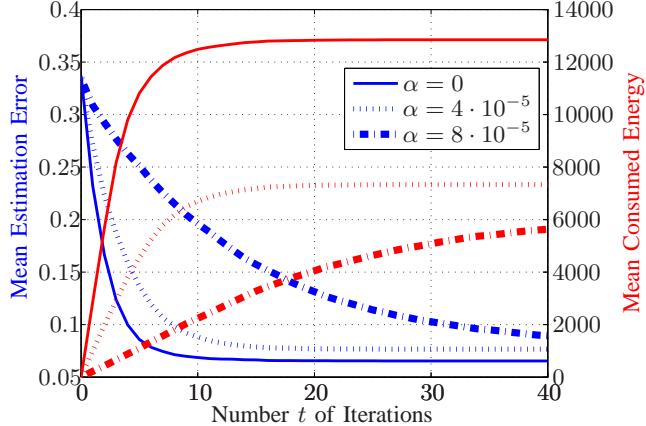}
 \caption{Proposed consensus algorithm for a WSN with $N=30$ nodes and small clusters $\lvert \mc{C}_i \rvert \subseteq \{2,\hdots,10\}$.}
\end{figure}

\begin{figure}[htb]
\label{fig:resultsnew2}
 \psfrag{Mean Estimation Error}{\textcolor{blue}{\small{Mean Estimation Error}}}
 \psfrag{Mean Consumed Energy}[t][c]{\textcolor{red}{\small{Mean Consumed Energy}}}
 \psfrag{Number t of Iterations}{\small{Number $t$ of Iterations}}
 \psfrag{0.0}{\small{$\alpha=0$}}
 \psfrag{8.8e-05              }{\small{$\alpha=8.8\cdot 10^{-5}$}}
 \psfrag{0.000156}{\small{$\alpha=1.6\cdot 10^{-4}$}}
 \psfrag{0}{\small{$0$}}
 \psfrag{0.1}{\small{$0.1$}}
 \psfrag{0.2}{\small{$0.2$}}
 \psfrag{0.3}{\small{$0.3$}}
 \psfrag{0.4}{\small{$0.4$}}
 \psfrag{5}{\small{$5$}}
 \psfrag{10}{\small{$10$}}
 \psfrag{15}{\small{$15$}}
 \psfrag{6}{\small{$6$}}
 \psfrag{8}{\small{$8$}}
 \psfrag{10}{\small{$10$}}
 \psfrag{12}{\small{$12$}}
 \psfrag{0.5}{\small{$0.5$}}
 \psfrag{1}{\small{$1$}}
 \psfrag{1.5}{\small{$1.5$}}
 \psfrag{2}{\small{$2$}}
  \psfrag{x 10}{\small{$\times 10^{4}$}}
 \psfrag{4}{\small{}}
 \psfrag{}{}
 \centering
  \includegraphics[width=8cm]{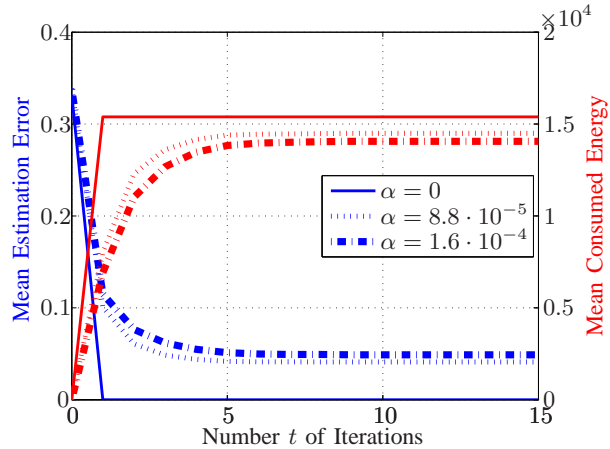}
 \caption{Proposed consensus algorithm for a WSN with $N=30$ nodes and large clusters $\lvert \mc{C}_i \rvert \subseteq \{20,\hdots,30\}$.}
\end{figure}

As can be seen from Fig. $2$ and $3$, the obtained simulation results confirm
the intended behaviour. In fact, if the regularization parameter is chosen to
be $\alpha=0$, the resulting consensus scheme is only tailored towards fast
convergence as in \eqref{equ:prob_only}, neglecting however the energy
consumption. Conversely, moving along the path of regularization parameter
results in a lower energy consumption at the cost of a slower convergence
rate. As an example, the results for a choice $\alpha\in\{4;8\}\cdot 10^{-5}$ in
the small cluster setting and $\alpha\in\{8.8;16\}\cdot 10^{-5}$ in the large
cluster setting are also depicted in Fig. $2$ and $3$, respectively.

\section{Conclusion}
In this paper, we presented a novel algorithm for the joint optimization of
convergence and energy consumption for consensus algorithms in wireless sensor
networks. The proposed algorithm takes into account distance dependent
transmit energies and clusters the network nodes according to user-defined
cluster sizes that may depend on application and site specific
characteristics. By incorporating a regularization term, we investigated
the trade-off between convergence speed and energy consumption. This is
achieved by a series of numerical simulations of a temperature monitoring
application under the assumption of noiseless communication links. The examples show
that the naive choice of a single cluster containing all nodes can be
outperformed in terms of network energy consumption if a certain excess error
can be tolerated.

Of particular interest for our subsequent work will be scenarios involving
noisy communications links and time-varying measurement objectives. This can
be achieved by using adaptive subgradient methods that harness interference
for computation \cite{CaSt13}. Also, by using more knowledge about the target
application and leveraging ideas from \emph{compressed sensing}, the number of
required measurements in the network can be reduced. In turn, this will result
in additional savings in energy consumption and introduce new degrees of
freedom to the tradeoff between convergence speed and estimation accuracy on
one hand, and energy consumption and network robustness on the other hand.




\section{Acknowledgments}
This work was supported by the German Research Foundation (DFG) 
under grants STA 864/3-2 and by the German Ministry for Education and Research
(BMBF) under grant 01BU1224.


\bibliographystyle{IEEEbib}
\bibliography{refs,myreferences}

\end{document}